# Even-Denominator Fractional Quantum Hall State in Conventional Triple-Gated Quantum Point Contact

Yasuaki Hayafuchi[1,‡], Ryota Konno[1,‡], Annisa Noorhidayati[1,‡], Mohammad Hamzah Fauzi[2], Naokazu Shibata[1], Katsushi Hashimoto[1,3,4], Yoshiro Hirayama[1,3,4,5*]

[1]*Graduate School of Science, Tohoku University, Sendai, Miyagi 980-8578, Japan*
[2]*Research Center for Physics, National Research and Innovation Agency, South Tangerang City, Banten 15314, Indonesia*
[3]*Center for Science and Innovation in Spintronics, Tohoku University, Sendai, Miyagi 980-8577, Japan*
[4]*Center for Spintronics Research Network, Tohoku University, Sendai, Miyagi 980-8577, Japan*
[5]*Takasaki Advanced Radiation Research Institute, QST, 1233 Watanuki-machi, Takasaki, 370-1292 Gunma, Japan*

*E-mail: yoshiro.hirayama.d6@tohoku.ac.jp

‡The first three authors contributed equally to this work.

The even-denominator states have attracted considerable attention owing to their possible applications in future quantum technologies. In this letter, we first report a 3/2 diagonal resistance, indicating the existence of a 3/2 state in a nanometer-sized triple-gated quantum point contact (QPC) fabricated on a high-mobility (not ultra-high-mobility) single-layer two-dimensional (2D) GaAs wafer. The center gate plays a crucial role in realizing the QPC's 3/2 state. Our observation of the 3/2 state using a conventional QPC device, which is a suitable building block for semiconductor quantum devices, paves a new path for the development of semiconductor-based quantum technologies.

Fractional quantum Hall states with even denominators have attracted attention owing to their novel carrier interactions and possible applications to error-free quantum manipulations. A well-known example is the $\nu = 5/2$ state where a non-Abelian topological order is expected.[1-3] To understand the background physics of the even-denominator fractional quantum Hall state and to also manipulate such a state in future quantum technologies, it is important to study the possibility of even-denominator states at the most fundamental ground Landau level, such as the $\nu = 3/2$ state. Despite the clear observation of the 5/2 state, the $\nu = 3/2$ state was not observed in a single-layer two-dimensional (2D) GaAs system even with an ultra-high mobility of $>4 \times 10^7$ cm$^2$/Vs.[4]

On the contrary, even-denominator fractional quantum Hall effects were observed at the ground Landau level ($\nu = 1/2, 3/2$) in certain systems, such as GaAs wide quantum wells with an effective bilayer carrier distribution,[5,6] GaAs bilayer systems,[7,8] ZnO/MgZnO-based quantum wells with a parallel field,[9] and bilayer graphene,[10] reflecting not only spin but also pseudospin and/or valley freedom characteristics. In a GaAs wide quantum well, both symmetric and antisymmetric states are used to reduce the Coulomb energy stabilizing the 331 state.[11,12] This state does not follow non-Abelian statistics; nevertheless, a change to the non-Abelian phase can be predicted by controlling parameters such as the tunnel coupling between the layers.[13,14]

Another interesting and important study reported the formation of the $\nu = 3/2$ state at the microscopic scale with an area of approximately 4 μm$^2$.[15,16] These observations were made using a single-layer 2D GaAs system (and not a bilayer system) with an ultra-high mobility of $>10^7$ cm$^2$/Vs, despite the lack of the $\nu = 3/2$ state in wide 2D systems starting from the same wafer. In possible applications of even-denominator non-Abelian systems to fault-tolerant quantum computation, the key issue is braiding; a small-area interference device may address this issue.[17-19] The clear observation of the $\nu = 3/2$ state in the microscopic regime is an interesting finding that can further the development of quantum information technology based on semiconductor microstructures.

On the other hand, the confined condition possibly modifies the state characteristics. Although thermal conductance[20] and NMR Knight-shift[21] measurements support the non-Abelian nature of the $\nu = 5/2$ state for a large sample, tunneling experiments have shown contradicting results for microstructures with the $\nu = 5/2$ state. Radu *et al.*[22] suggested a non-Abelian state; however, similar experiments conducted by Lin *et al*.[23] and Baer *et al*.[24] support an Abelian 331 state. Notably, recent experiments conducted by Fu *et al*.[25] suggested that the potential shape of the confinement plays an important role. The obtained characteristics favor a non-Abelian state for a soft confinement at a relatively high temperature.[25] Moreover, in a microstructure, we can expect a complicated edge channel scheme, including counter-propagating channels, which are hidden in large-scale samples.[26-29]

These results have motivated us to test the possibility of a 3/2 state formation in a conventional well-established quantum point contact (QPC) that can flexibly control the shape of one-dimensional (1D) confinement potential. Our experiments raise questions on the need for ultra-high-mobility wafers. QPCs are the most important building blocks of future semiconductor devices in quantum technology; therefore, our study has important implications for advancements in semiconductor-based quantum information technology.

In this study, we experimentally investigated the transport characteristics of a triple-gated QPC with a center gate in the fractional quantum Hall regime. The fabricated QPC structure, schematically shown in Fig. 1(a) along with its scanning electron microscopy (SEM) image (Fig. 1(b)), has a nominal length $L$ of 400 nm, a separation between the split Schottky gates $W$ of 600 nm, and a 200-nm-wide center gate. In addition to the split Schottky gates and center gate, this device has a back gate. We can control the 2D electron density in the 20 nm GaAs quantum well sandwiched by $Al_{0.33}Ga_{0.67}As$ barriers using the back gate voltage ($V_{bg}$). The quantum well is located 175 nm below the surface. The low-temperature electron mobility is $\mu = 1.47 \times 10^6$ cm$^2$/Vs at electron density of $1.8 \times 10^{11}$ cm$^{-2}$.

The transport characteristics of the triple-gated QPC were first measured under a zero magnetic field. Figure 1(c) shows the results. We observed quantized conductance at a temperature of 100 mK, particularly when a positive $V_{cg}$ (center gate voltage) was applied, reflecting the successful fabrication of the QPC device. As expected for a triple-gated device, the conductance curve gradually shifts to the negative $V_{sg}$ (voltage applied to the split Schottky gates) region with increasing $V_{cg}$.[30] Here, we applied the same bias $V_{sg}$ to the pair of split gates. The sharp increase in the conductance at approximately $V_{sg} = -0.3$ V corresponds to the transition from 1D to 2D. The $V_{sg}$ value to deplete the 2D electrons under the split gate depends only on the 2D electron density determined by $V_{bg}$; hence, the transition point becomes insensitive to $V_{cg}$.

Next, we applied a magnetic field $B = 6$ or 7 T and measured the characteristics in the fractional quantum Hall regime. We measured the diagonal resistance ($R_{diag} = V_D/I_{AC}$) as schematically shown in Fig. 1(a). Although a magneto-depopulation of the 1D sub-band increases the plateau length,[31] there is a gradual shift in the $R_{diag}$ curve with the center gate voltage $V_{cg}$, similar to the zero-field characteristics shown in Fig. 1(c), when the filling factor of the bulk region, $\nu_{bulk}$, is not 5/3. The situation changes when $\nu_{bulk}$ is set at 5/3, as shown in Fig. 2(a). When $V_{cg}$ is increased from −0.1 to 0.1 V, the $R_{diag}$ curve gradually shifts, and the $R_{diag}$ value decreases with $V_{cg}$, as expected. However, with further increase in $V_{cg}$, $R_{diag}$ starts to increase and finally stabilizes at 17.2 kΩ, which corresponds to the $\nu_{QPC} = 3/2$ value. This behavior can be clearly confirmed in Fig. 2(b), which represents $V_{cg}$ dependence of $R_{diag}$ at $V_{sg} = -1.05$ V and $B = 6$ T. This is a unique behavior and the main experimental finding of this study. Once decreased, $R_{diag}$ starts to increase with $V_{cg}$ despite the fact that a positive $V_{cg}$ supplies electrons in the QPC channel. This suggests that the inside of the QPC is maintained at $\nu_{QPC} = 3/2$. The bulk region on both sides of the QPC is maintained at $\nu_{bulk} = 5/3$ in this experiment; hence, the diagonal resistance corresponding to $\nu_{QPC} = 3/2$ suggests $\nu = 1/6$ scattering channels at both edges of the QPC, as discussed by Fu et al..[16] The expected edge channel schematic diagram near the QPC center region is discussed in the supplementary information S1.

We conducted similar measurements using conventional QPCs with the same nominal width of $W = 600$ nm but without the center gate to prove the importance of the center gate (see Supplementary Information S2). The obtained results indicate that the lateral spread of the depletion from the side gates is too high, and therefore, the electron density in the QPC becomes lower than that at $\nu_{QPC} = 3/2$ for the rather narrow channel with a width of 600 nm. This may explain why a wider (approximately 2 μm) confinement was required to observe the $\nu_{QPC} = 3/2$ structure in previous experiments,[15,16] where there were no center gates.

A comparison between the two devices (with and without the center gate) confirms the essential role of the center gate in forming the $\nu = 3/2$ structure in the conventional QPC whose width is <1 μm. A positive $V_{cg}$ increases the electron density in the narrow channel, reaching $\nu_{QPC} = 3/2$ even for a channel width of 600 nm. The center gate also contributes to effectively suppressing the disorder potential. In our previous QPC experiments using a medium-mobility wafer ($\mu = 3\times10^5$ cm$^2$/Vs at $n = 2\times10^{11}$ cm$^{-2}$), clear quantized conductance could be observed under the application of positive $V_{cg}$; otherwise, there was no quantized conductance due to the influence of strong potential disorder.[32] This probably suggests another important role of the positive center gate voltage, i.e., effective screening of the disorder potentials. This can partly explain why the even-denominator fractional quantum Hall states were observed in QPCs fabricated on high-mobility 2D systems and did not require ultra-high-mobility systems.

Figure 3(a) plots the regions where $R_{diag}$ becomes the value of ±0.8% from the exact $2/3\times(h/e^2)$ resistance. The regions are indicated in blue. These data clearly show that the $\nu_{QPC}$ = 3/2 plateau appears under conditions where $V_{sg}$ is sufficiently negative to deplete the 2D electrons under the split gate and form a constricted region, and where $V_{cg}$ is sufficiently positive to supply electrons in the constricted region to maintain the $\nu_{QPC}$ = 3/2 state. The obtained results suggest that electrons are redistributed in the microscopic QPC area to maintain the QPC state at $\nu_{QPC}$ = 3/2 when the center gate can supply sufficient electrons to the constricted area. At the highest $V_{cg}$ of 0.65 V, the total electron number becomes excessive to ensure a stable $\nu$ = 3/2 area, showing small but observable fluctuations.

Furthermore, we experimentally verified how the plateau value varies with the experimental parameters, such as $V_{bg}$ and magnetic field. Figure 3(b) shows the appearance of the plateau regions under varying $V_{bg}$. As shown in the inset of Fig. 2(a), $R_{xx}$ becomes minimum almost at the center of the $\nu_{bulk}$ = 5/3 plateau in $R_{xy}$. On both sides, $R_{xx}$, i.e., the bulk resistance, gradually increases, particularly in the higher $V_{bg}$ side. Notably, the $R_{diag}$ value reflects this series resistance change in the $\nu_{bulk}$ = 5/3 region, as shown in Fig. 3(c). Despite such an effect of the series resistance, the $V_{sg}$ range in which the $\nu_{QPC}$ = 3/2 plateau appears is almost constant in the $V_{bg}$ region of 1.58–1.64 V (Fig. 3(b)), independent of the series resistance of the bulk, namely backscattering in the bulk region. This result suggests the robust feature of the $\nu_{QPC}$ = 3/2 plateau. A similar behavior can be observed when the magnetic field is changed, as shown in Figs. 3(d) and (e). The electron density corresponding to the $\nu_{bulk}$ = 5/3 decreases with the magnetic field, with the threshold $V_{sg}$ value for the formation of the constricted region becoming less negative. Correspondingly, the QPC 3/2 plateau region also moves to less negative $V_{sg}$ with decreasing magnetic field (Fig. 3(d)).

The $\nu_{bulk}$ = 5/3 fractional state becomes less pronounced with decreasing magnetic field, and the series resistance increases, as shown in Fig. 3(e). Finally, the plateau value largely deviates from the exact 3/2 value at $B$ = 4 T. Nevertheless, the appearance of the wide plateau region in Fig. 3(d) once again demonstrates the robustness of the formation of the QPC 3/2 state. When we increase the measurement temperature, the bulk 5/3 state becomes obscure and the $\nu_{QPC}$ = 3/2 plateau disappears similar to the low magnetic field case.

Here, the accuracy of the plateau quantization should be considered. In previous reports,[15,16] the plateau resistance was reported to be within ±0.02% of the exact 2/3×($h/e^2$) resistance for the approximately 4 μm$^2$ constricted region when using an ultra-high mobility 2D system at a very low temperature of approximately 20 mK. A precise quantization is the essence of the quantum Hall effect. However, as discussed, the plateau values are affected by the series resistance in the bulk areas. When not using ultra-high mobility 2D systems, the $R_{xx}$ value of the bulk area is not completely zero; hence, we cannot expect extremely accurate quantization in our experimental setup. Although we cannot expect a good quantization accuracy, our ±0.8% accurate plateau value and experimentally obtained characteristics clearly suggest the existence of the $\nu_{QPC}$ = 3/2 state not in ultra-high-mobility but in high-mobility QPC with the conventional size of < 1 μm.

The origin of the $\nu_{QPC}$ = 3/2 state is unclear. Although the 20 nm quantum well used in the present experiment excludes the possibility of bilayer formation, unlike wide wells,[5,6] a complicated edge channel in the microscopic structure may allow the formation of the local hole edge channel and stabilize the $\nu_{QPC}$ = 3/2 state by the similar mechanism with the 331 state. Theoretical studies have indicated an important role of the finite quantum well thickness in producing the incompressible fractional quantum Hall state, particularly the $\nu$ = 5/2 state.[33,34] The modification of the vertical electron distribution by the front and back gates might play a similar role as the quantum well thickness. It also contributes to a change in the edge characteristics. Whether the $\nu_{QPC}$ = 3/2 state is Abelian or non-Abelian is a sensitive question; nevertheless, the flexible tunability of the confinement potential of the triple-gated QPC with a back gate may allow switching between the Abelian and non-Abelian states.[13,25] Further study is required to understand the role of spin freedom.[35] Tunneling experiments through the constriction[22-25] and resistively-detected nuclear-magnetic resonance[21,36,37] can help clarify the detailed characteristics of the $\nu_{QPC}$ = 3/2 state.

In conclusion, we have demonstrated appearance of the even-denominator fractional quantum state in the conventional triple-gated QPC device. We found the plateau feature corresponding to the $\nu_{QPC} = 3/2$ state in the diagonal resistance when the bulk filling sets to $\nu_{bulk} = 5/3$ and the center-gate voltage is positive enough to keep electron density in the center region of the QPC at the 3/2 filling. Another important point of our finding is that we can see the QPC 3/2 state by using conventional high-mobility but not ultra-high-mobility device thanks to the screening effect of the center gate. Although further studies are necessary, our finding suggests a possible formation of the unique even-denominator fractional-quantum-Hall state in the nanoscale local area by conventional high-quality two-dimensional systems that can be easily achieved with modern crystal growth technology. It will pave a way to open new quantum information devices based on semiconductor quantum systems. It will also give us a hint to consider edge channel formation in the fractional-quantum-Hall regime.

**Acknowledgments**

We thank K. Muraki and NTT for supplying high-quality GaAs 2D wafers, and M. Takahashi, M. F. Sahdan, K. Nagase, and K. Sato for their help in fabricating the device. K.H. acknowledges JSPS for financial support (KAKENHI 26390006, 17H02728). Y. Hi. acknowledges support from JSPS (KAKENHI 15H0587, 15K217270, and 18H01811). M.H.F., K.H., and Y. Hi. thank Tohoku University's GP-Spin program for their support.

**Figure Captions**

**Fig. 1.** (a) Schematic of the fabricated triple-gated QPC device. (b) SEM image of the center region of the QPC. (c) QPC conductance as a function of $V_{sg}$ measured at 100 mK under a zero magnetic field. The $V_{cg}$ is changed from 0 to 0.5 V with a step of 0.05 V.

**Fig. 2.** (a) $R_{diag}$ through the QPC as a function of $V_{sg}$ measured at $B = 7$ T. In this measurement, the bulk filling of both sides of the QPC is set to $\nu_{bulk} = 5/3$, which was realized at $V_{bg} = 2.18$ V as confirmed by the $R_{xy}$ data shown in the inset (The data in the inset were measured at $V_{sg} = 0.3$ V and $V_{cg} = 0$ V). With increasing $V_{cg}$ up to 0.1 V, the $R_{diag}$ value decreases, and the curve shifts to more negative $V_{sg}$, as expected. However, with further increase in $V_{cg}$, $R_{diag}$ increases and reaches the 3/2 resistance value, suggesting the formation of a novel $\nu_{QPC} = 3/2$ state. (b) Similar features can be observed at 6 T as shown in the inset. The main figure indicates how $R_{diag}$ varies as a function of $V_{cg}$, where $V_{sg} = -1.05$ V and $V_{bg} = 1.62$ V corresponding to the $\nu_{bulk} = 5/3$. The $R_{diag}$ value once decreases below the 3/2 resistance value but increases and reaches the plateau at the 3/2 resistance value with further increase in $V_{cg}$. All the measurements were made at 100 mK.

**Fig. 3.** (a) $V_{cg}$ dependence of the 3/2 plateau area as a function of $V_{sg}$. $V_{bg}$ is set to 1.62 V corresponding to the $\nu_{bulk} = 5/3$ at $B = 6$ T. The blue color indicates the regions where $R_{diag}$ becomes ±0.8% of the 3/2 resistance. The vertical solid line indicates the transition from 2D to 1D. (b) $R_{diag}$ regions with ±0.8% of the 3/2 resistance appear in the $V_{bg}$ regions corresponding to the bulk filling at approximately $\nu_{bulk} = 5/3$ ($B = 6$ T). (c) Plateau resistance as a function $V_{bg}$. The deviation in the bulk filling from the exact 5/3 increases the plateau value, suggesting a contribution of the series resistance in the bulk region. (d) Variation in the $R_{diag}$ regions with ±0.8% of the 3/2 resistance with the magnetic field ($B$). Under all the magnetic fields, $V_{bg}$ is set to the value corresponding to the $\nu_{bulk} = 5/3$ as shown in (e). All the measurements were performed at 100 mK.

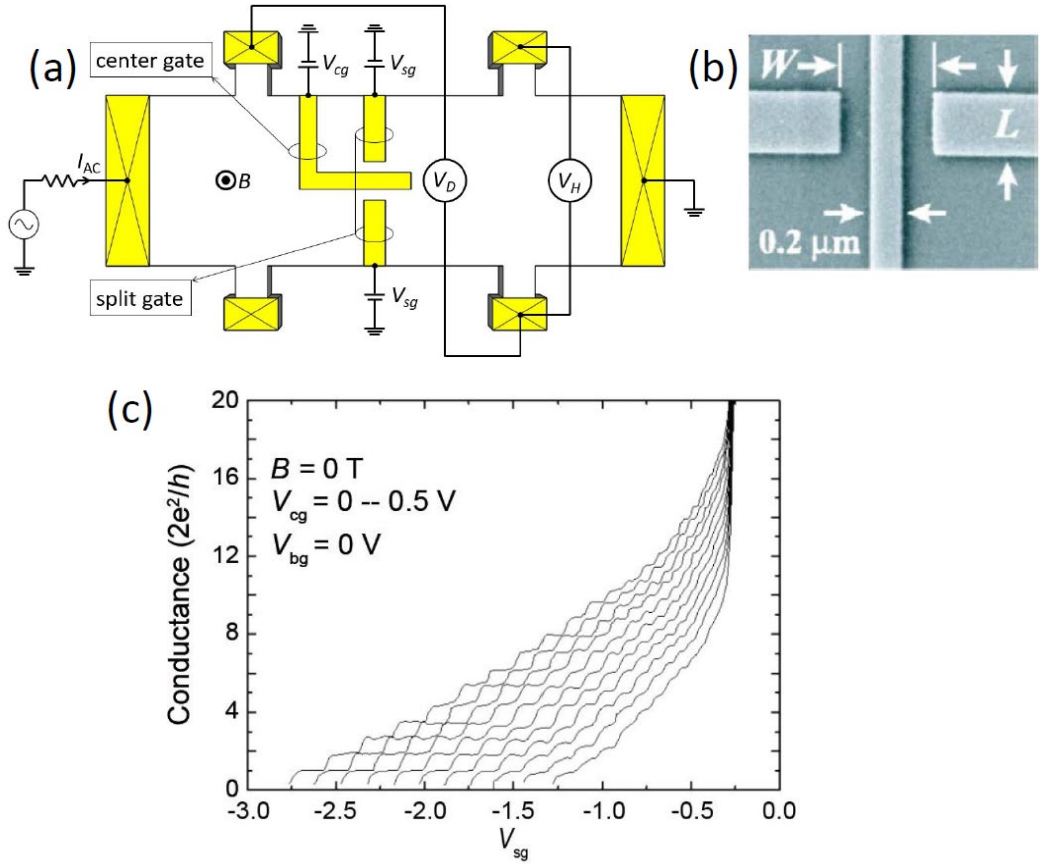

**Fig. 1**

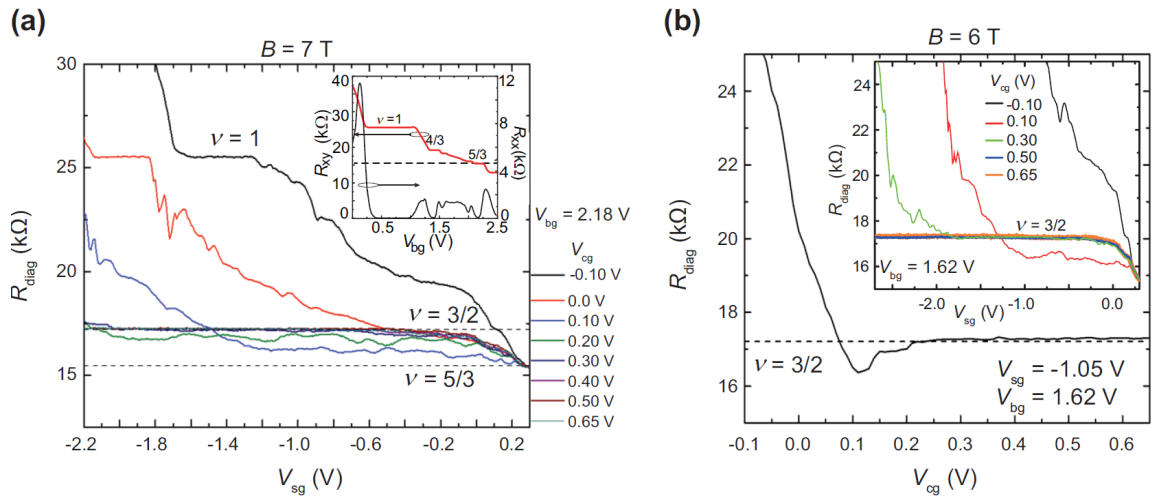

**Fig. 2**

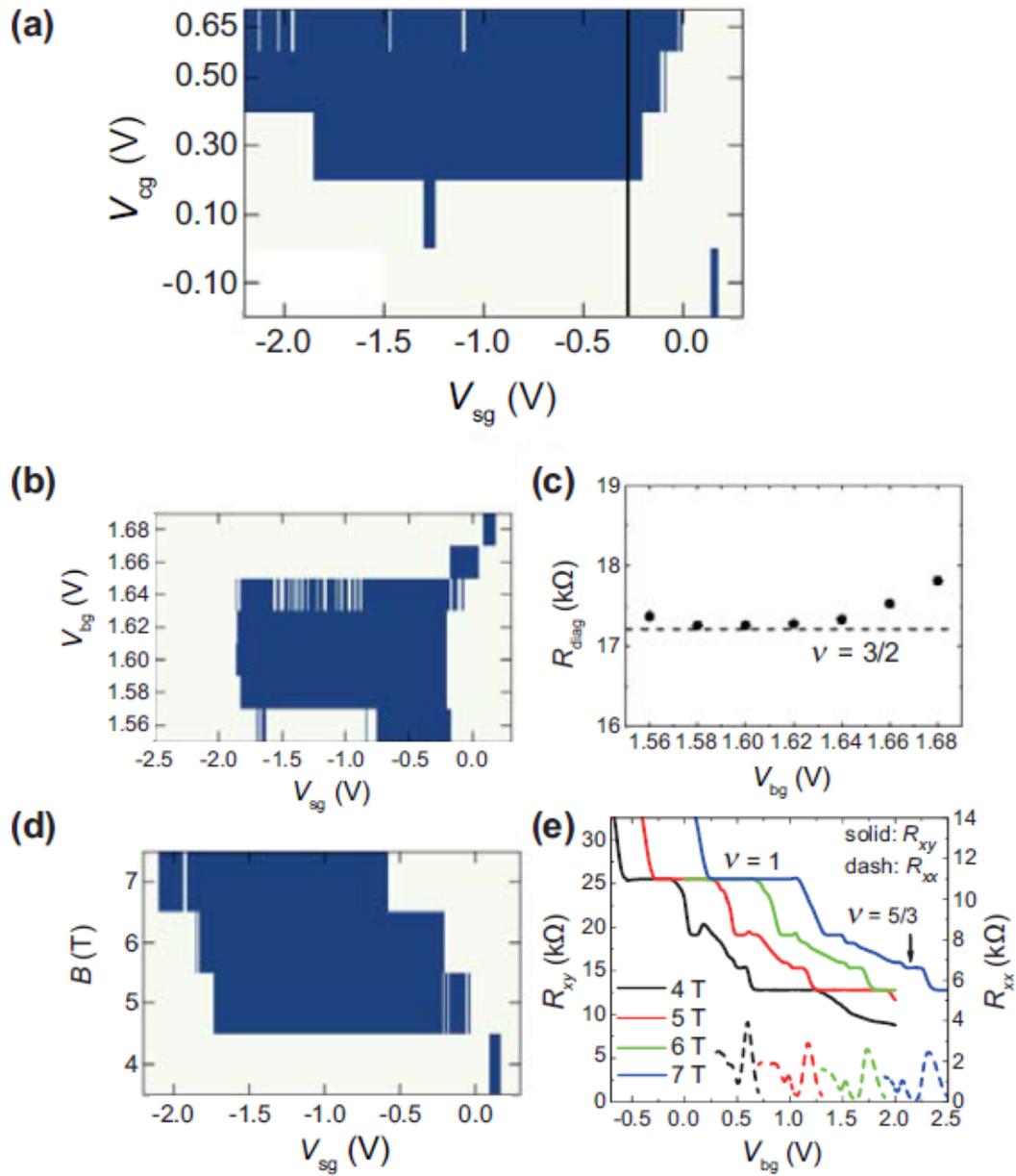

**Fig. 3**

# Supplementary information: Even-Denominator Fractional Quantum Hall State in Conventional Triple-Gated Quantum Point Contact

Yasuaki Hayafuchi[1,‡], Ryota Konno[1,‡], Annisa Noorhidayati[1,‡], Mohammad Hamzah Fauzi[2], Naokazu Shibata[1], Katsushi Hashimoto[1,3,4], Yoshiro Hirayama[1,3,4,5*]

[1]*Graduate School of Science, Tohoku University, Sendai, Miyagi 980-8578, Japan*
[2]*Research Center for Physics, National Research and Innovation Agency, South Tangerang City, Banten 15314, Indonesia*
[3]*Center for Science and Innovation in Spintronics, Tohoku University, Sendai, Miyagi 980-8577, Japan*
[4]*Center for Spintronics Research Network, Tohoku University, Sendai, Miyagi 980-8577, Japan*
[5]*Takasaki Advanced Radiation Research Institute, QST, 1233 Watanuki-machi, Takasaki, 370-1292 Gunma, Japan*

*E-mail: yoshiro.hirayama.d6@tohoku.ac.jp

‡The first three authors contributed equally to this work.

**S1. Schematic diagram near QPC at $\nu_{QPC} = 3/2$**

The physics behind the formation of $\nu_{QPC} = 3/2$ structure is still unclear yet so that it is difficult to draw an exact edge channel diagram. Nevertheless, we show in Fig. S1 an expected edge channel schematic diagram near the QPC center region when $\nu_{QPC} = 3/2$ is achieved by setting the bulk region to $\nu = 5/3$, full depletion under the pair of split gates, and applying an appropriate positive bias, $V_{cg}$, to the center gate. For the sake of clarity, the notation with a unified denominator of 6 is also shown. In this figure, the medium blue areas show bulk 5/3 state. The white areas show depleted regions, which laterally expand from the split gate area, except for the area between the side gates where the electron density is increased by positively biased center gate. The light-blue area shows the most important $\nu_{QPC} = 3/2$ state. The robust feature of the $\nu_{QPC} = 3/2$ state to $V_{cg}$ suggests that the $\nu_{QPC} = 3/2$ state is maintained at its stable condition by adjusting the electron distribution in the constricted region. The dark-blue color areas may have a density higher than $\nu = 5/3$ due to the positive $V_{cg}$. Although they have no direct effect and are not shown here, there may be circulating edge channels inside the higher-density regions. We can expect the existence of 1/6 tunneling channels in the connecting regions at both sides of the QPC. The diagonal 5/3 regions operate as the ohmic contacts to the $\nu_{QPC} = 3/2$ state in the QPC region. This makes the diagonal resistance of 17.2 kΩ corresponding to the $\nu_{QPC} = 3/2$ state similar to the diagonal resistance observed in the quantum Hall plateau.

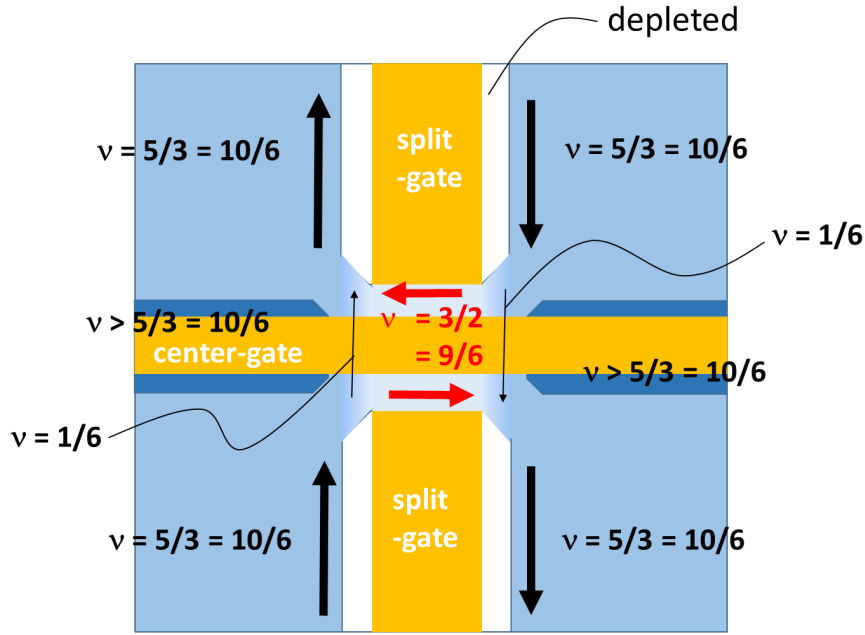

**Fig. S1**. Schematic diagram of edge channel around QPC center region when $\nu_{QPC} = 3/2$ is achieved. Arrows indicate the expected edge channels in the fractional-quantum-Hall regime and 1/6 tunneling. The arrow direction is reversed by magnetic field reversal.

**S2. Characteristics of QPC without center gate**

To prove the important role of the center gate in defining the $\nu_{QPC} = 3/2$ state, we conducted similar measurements using conventional QPCs without the center gate. This device has a gap of 600 nm between the split gates, as schematically shown in Fig. S2 (a), and is fabricated on the same wafer. The successful fabrication of the QPC device is confirmed by the clear quantized conductance step observed at $B = 0$ T (Fig. S2 (b)). The quantized step becomes prominent with increasing $V_{bg}$, hence electron density of the two-dimensional system. In the next step, we set the device at 7 T, tune the bulk filling at $\nu_{bulk} = 5/3$ by controlling $V_{bg}$, and then measure $R_{diag}$ as a function of $V_{sg}$. Figure S2 (c) shows the obtained result. There is no observation at $R_{diag}$ corresponding to $\nu_{QPC} = 3/2$. From the change in the QPC conductance slope observed at $B = 0$ T (Fig. S2(b)), we can estimate the $V_{sg}$ value at which the electrons under the split gates are fully depleted, and the center of the QPC changes to the constricted wire region. When $V_{bg} = 3.54$ V, the wire confinement is expected in the region of $V_{sg} < -0.75$ V. At this point, $R_{diag}$ increases beyond the value expected at $\nu_{QPC} = 3/2$ (17.2 k$\Omega$). This result indicates that the lateral spread of the depletion from the side gates is too high, and therefore, the electron density in the QPC becomes lower than that at $\nu = 3/2$ for the rather narrow channel with a width of 600 nm. The situation was the same when we changed $V_{bg}$ and realized $\nu_{bulk} = 5/3$ at different magnetic field.

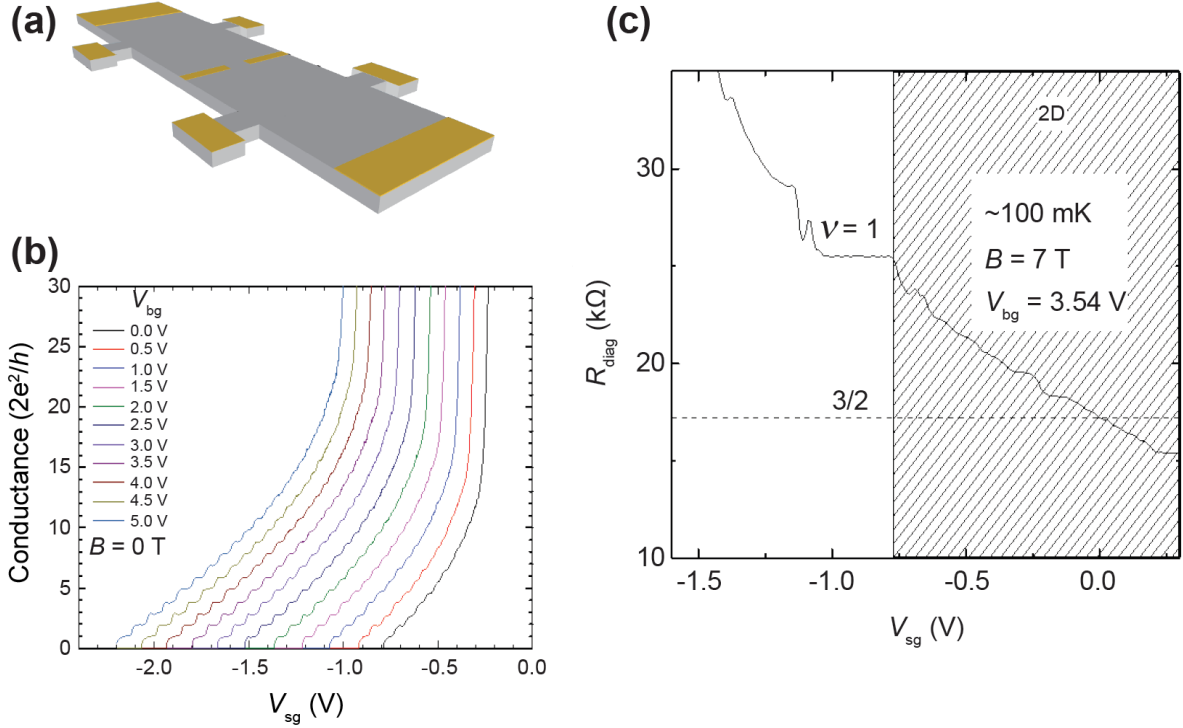

Fig. S2. (a) Schematic of the QPC device without the center gate. (b) Zero magnetic field QPC conductance as a function of $V_{sg}$, where the same $V_{sg}$ is applied to both the side gates. The parameter, $V_{bg}$, changes the 2D bulk electron density. A clear quantized conductance is observed in the entire $V_{bg}$ region, reflecting a high-quality QPC device. (c) Example of $R_{diag}$ as a function of $V_{sg}$ at $B = 7$ T. The filling of the bulk region is set to $\nu_{bulk} = 5/3$ by applying $V_{bg} = 3.54$ V to the back gate. The shaded area indicates the 2D region where depletion under the split gates is incomplete, judging from the 1D-to-2D transition estimated from (b). For the 600-nm-wide QPC without the center gate, the electron density in the center region of the QPC is always lower than the density corresponding to the 3/2 filling even when the magnetic field (then $V_{bg}$) is changed. All measurements were performed at 100 mK.